\documentstyle[12pt]{article}
\normalsize
\begin{document}
\overfullrule 0 pt 
\null
\begin{center} 
{\Large {\bf Kinetic 
and transport equations for localized excitations in Sine-Gordon
model}}
\vspace{1cm}
{\bf \large 
\vskip 0.1 in 
I.V. Baryakhtar \\
{\it B.I. Verkin Institute for Low Temperature Physics and Engineering, Lenin 
Ave. 47, Kharkov 310164, Ukraine}
\vskip 0.05 in
V.G. Baryakhtar \\
{\it Institute of Magnetism, 36b Vernadsky St., Kiev 252142, Ukraine}
\vskip 0.05 in
E.N.Economou\\
{\it Research Center of Crete, FORTH, P.O. Box 1527, 71110 Heraklion, Crete,
Greece}}
\end{center}
\vskip 0.2 in


\begin{center} {\bf Abstract} \end{center}

We analyze the kinetic behavior of localized excitations
- solitons, breathers and phonons - in Sine-Gordon model. 
Collision integrals for all type of localized excitation
collision processes are constructed, and the kinetic equations are derived. 
We prove that the entropy production in the system of localized
excitations takes place only in the case of inhomogeneous distribution of
these excitations in real and phase spaces. We derive transport equations 
for soliton and breather densities, temperatures and mean velocities i.e. 
show that collisions of localized excitations lead to 
creation of diffusion, thermoconductivity and intrinsic friction processes.
The diffusion coefficients for solitons and breathers, describing the
diffusion processes in real and phase spaces, are calculated. It is
shown that diffusion processes in real space are much faster than the
diffusion processes in phase space.

\newpage
\vskip 0.1 in
\begin{center}
{1. INTRODUCTION}
\end{center}

The problem of kinetic properties of excitations in integrable models
belongs to a class of most nontrivial problems of physical kinetics. First,
enormously long relaxation of nonlinear excitation has been found in
numerical experiments by Fermi Pasta and Ulam [1]. In Zabusky and Kruskal
numerical experiment [2] unexpected behavior of localized nonlinear
excitations has been discovered, namely their interaction without changing
their forms and velocities. Zabusky and Kruskal named them solitons. Shortly
after, the analytical method of solving nonlinear differential equation with
partial derivatives - inverse scattering method - was found in the framework
of Korteweg - de Vries equation for 1+1 dimensions [3]. Other physically
reasonable continuous models, treatable by the inverse scattering method are
the Nonlinear Schr\"odinger (NS) equation, the Sine-Gordon (SG) equation and
the Landau-Lifshitz (LL) equation [4,5]. Common excitations in integrable 
models are: 
the one parameter localized wave i.e. the soliton and the nonlinear
periodic wave for KdV model; and the two parameter localized wave i.e. the
breather for SG and LL models; and the envelope soliton for NS equation. All
these excitations interact without changing their forms and their velocities
and therefore their energies; the only result of their interaction are
shifts of their coordinates and for breathers and NS solitons change of
their phases as well. It is worthwhile to emphasize that many particle
effects are absent in the following meaning: the total shift in collisions 
involving several excitations is equal to the sum of shifts in each collision 
(see[4]).

Soon after analytically solving the KdV equation [3], Zakharov considered 
the possibility of the kinetic decryption of 
solitons [6]. In [6], kinetic equation for solitons was
written but only the effect of renormalization of soliton velocity due to
soliton - soliton collisions has been considered. Another approach for the
investigation of kinetic properties of kink type solitons was proposed in
[7-9]. In [7] numerically and in [8,9] analytically the diffusion of kinks
interacting with phonons in the $\phi^4$ model was considered. This model is
not exactly integrable, but the potential for kink - phonons interaction has
reflectionless form and the result of the interaction is the same as for
soliton - soliton collision in integrable models. Taking into account such
kind of interaction in [8,9], kink diffusion coefficient was calculated. A bit
later the same calculations have been done for SG model [10].
The renormalization of soliton velocity and the diffusion coefficient of the 
solitons due to soliton - soliton  and soliton - magnon collisions have been 
considered  for SG model in [11]. 

To explain the effect of velocity renormalization 
it is useful to note that the dependence of soliton coordinate $x$ from time 
can be written in the following form:
$$x(t)=x_0 + v_1 t+\int\limits_{-\infty }^tdt^{\prime }\int dx_2dv_2
|v_1-v_2|  \Delta
x_s\left( v_1,v_2\right) \theta \left( t-t^{\prime }\right)
f(x_2,v_2,t^{\prime }), \eqno(1.1) $$
when shifts of soliton coordinates $\Delta x_s(v_1,v_2)$ are taken into 
consideration. In
formula (1.1) $f(x,v,t)$ is the distribution function of solitons or phonons
(magnons), 
and the sample soliton collides with soliton  or phonon (magnon) wave packet
$(x_2,v_2)$ at the moment $t'$,

$$\theta \left( t\right) =1,{~~}t>0,{~~~~~~}\theta \left(t\right)=0,{~~}t<0.
\eqno (1.2)$$

>From formula (1.1) one can obtain the following expression for the average
velocity of the sample soliton:
$$< v_1> =v_1+\int dx_{2} dv_{2} 
|v_1-v_2| \Delta x_{s} (v_{1},v_{2} ) f(x_{2} ,v_{2} ,t) \eqno(1.3) $$
The diffusion coefficient of soliton is give by:
$$ D= {{<[x (t)-<v> t -x(0)] ^2>} \over 2 t}, \eqno(1.4)$$
(see [8-10]).

So the shift of soliton coordinate $\Delta x$ leads  to its diffusion. Do 
another kinetic coefficients for solitons exist in the integrable models?
This question is closely related with the problem of entropy production.
>From the fact of conserving of the solitons 
velocities follows the absence of the  entropy production in 
momentum space only. But in the coordinate space the interaction of solitons 
is very strong. Therefore the question: " Does entropy increase 
in integrable systems?"- is not a nonsense. The answer on these questions 
have been done in [12] on the example of kink-type solitons in SG model.
In [12] kinetic equation for the gas of kinks has been formulated, the entropy
production has been proved and the existence of the coefficients of
intrinsic friction and thermal conductivity have been shown.    
Obviously that  such kinetic coefficients as mobility coefficient appear in
the presence of perturbations, which destroy the integrability of the model
( see [13 ] and review [14] ).

In this paper the kinetic behavior of solitons, breathers and phonons in
the framework of the SG model is considered. We  construct collision
integrals for all possible collisions and formulate the system of Boltzmann
type kinetic equations for solitons, breathers and phonons. Based on the
system of kinetic equations thus obtained, we  prove that the entropy
production takes place as a result of randomization of the distribution of
excitations in coordinate and phase space. Thus we are able to derive 
transport equations for solitons and breathers and to calculate self-diffusion
coefficients for soliton-soliton and soliton-breather and breather-breather 
collisions as an example.
\vskip 0.1 in
\begin{center}
{2.ELEMENTARY EXCITATIONS IN SINE -GORDON EQUATION}
\end{center}
\vskip 0.1 in
Sine-Gordon equation in dimensionless variables can be written as: 
$$
\varphi _{tt}-\varphi _{xx}+\frac{m^2}\beta \sin \beta \varphi =0. \eqno(2.1) 
$$
Here $\varphi $ is the order parameter, t is the dimensionless time, x is
the dimensionless space coordinate, c=1 is the characteristic velocity, m is
the dimensionless mass and $\beta $ is the parameter of nonlinearity.

Eq.(2.1) follows from Hamilton's equations (see, for example, [5]): 
$$
\varphi _t=\left\{ H,\varphi \right\} ,\pi _t=\left\{ H,\pi \right\} \eqno(2.2) 
$$
with Hamiltonian: \vskip 0.02 in 
$$
H=\frac 12\int \left[ \pi ^2+\varphi _x^2+2\left( \frac m\beta \right)
^2\left( 1-\cos \beta \varphi \right) \right] dx\equiv \int h\left( \varphi
,\pi \right). \eqno(2.3) 
$$
In eqs. (2.2) -(2.3) $\varphi $ and $\pi $- are canonically conjugated
coordinate and momentum: 
$$
\left\{ \pi \left( x\right) ,\varphi \left( y\right) \right\} =\delta \left
(x-y\right),                     \eqno(2.4) 
$$
where the Poisson brackets are defined by usual way: 
$$
\left\{ A,B\right\} =\int \left( \frac{\delta A}{\delta \pi (x)}\frac{\delta
B}{\delta \varphi (x)}-\frac{\delta A}{\delta \varphi (x)}\frac{\delta B}{
\delta \pi (x)}\right) dx. \eqno (2.5) 
$$
It is easy to see that the functionals of total momentum, $P$, of the system: 
\vskip 0.1 in 
$$
P=-\int \pi \left( x,t\right) \frac{\partial \varphi \left( x,t\right) }{
\partial x}dx \eqno(2.6) 
$$
and total ''angular'' momentum K: \vskip 0.1 in 
$$
K=\int xh\left[ \pi ,\varphi \right] dx \eqno(2.7) 
$$
commute with Hamiltonian (2.3).

It is well known that there are two types of localized excitations (LE) in SG
model: solitons and breathers. Soliton is a one parameter solution, breather
is a two parameter solution, the breather can be interpreted as bound states
of two solitons (see [4,5]). The one parameter solution of
the SG model, often named as kink, can be written as: \vskip 0.05 in 
$$
\varphi _s =-4\frac \epsilon \beta arctg\left[ \exp \left( \delta
_s^{-1}(x-v_st-x_{0s}\right) \right] .\eqno(2.8)
$$
Breather solution has the following form: \vskip 0.1 in
$$
\varphi _b=\frac 4\beta  arctg \left( \frac{\omega _2}{\omega _1}
\right) \frac{sin\left[ \omega (v_b)t-k(v_b)x-\varphi _{0b}\right] }{
ch\left[ \left( \delta _b^{-1}(x-v_bt-x_{0b}\right) \right] }, 
\eqno(2.9)
$$
where: 
$$
m(v)=m/\sqrt{1-v^2},{~~~}\omega (v)=m(v)\omega _1,{~~~}k(v)=vm(v)\omega _1, 
$$
$$
\delta _s^{-1}=m(v_s),{~~~~}\delta _b^{-1}=\omega _2m(v_b)\quad \quad. \eqno
(2.10) 
$$
Here $\epsilon $=$\{$+1,-1,0$\}$ is the soliton topological charge, +1
corresponds to soliton, -1 to antisoliton, 0 to breather, $v_s$ and $v_b$
are the soliton and breather velocities, $\delta_s$ and $\delta _b$
are the soliton and breather sizes, $x_{0s}$ and $x_{0b}$ are the initial
coordinates of soliton and breather, $\varphi _{0b}$ is the initial
breather phase, $\omega_1$ and $\omega _2$ are the breather parameters
satisfying the following condition (see [5]): 
$$
\omega _1^2+\omega _2^2=1. \eqno(2.11) 
$$
When $\omega _2\rightarrow 0$, the breather solution reduces to the following
ones:
$$
\lim_{\omega _2\rightarrow 0}\varphi _b(x,t) =\varphi _p(x,t)=4{\frac{
\omega _2}\beta }{\frac{sin[\omega (v)t-k(v)x-\varphi _{0b}]}{ch[m(v) \omega_2
(x-kt/\omega -x_{0b})]}},\eqno(2.12) 
$$
and  gives a plane wave solution i.e. a phonon.

The phonon frequency is related with wave vector {\bf k} by formula:
$$
\omega _{ph}^2\left( k\right) =m^2+k^2. \eqno(2.13) 
$$
The breather velocity, when $\omega _2 \rightarrow 0$, reduces to the phonon
group velocity:
$$
v_{ph}=\left( k/\omega _{ph}\right) = \left( \partial \omega _{ph}/\partial
k\right), \eqno(2.14) $$
and $4\omega $ $_2$ becomes the amplitude of phonon oscillations.

>From formulas (2.8), (2.9) and (2.3) it is easy to find the soliton and the
breather energy:
$$
E_s=M_s/\sqrt{1-v_s^2},{~~~}E_b=M_b/\sqrt{1-v_b^2}. \eqno(2.15) 
$$
The soliton and the breather rest masses are given by:
$$M_s=8^{}m,{~~~} M_b = 16m \omega _2. \eqno(2.16)$$
When $\omega _2\rightarrow 1$, the breather reduces to soliton-antisoliton
bound state with a mass $16m$. The difference between the breather energy on
and the sum of energies of isolated soliton  and antisoliton gives
the binding energy of the breather:
$$
\Delta E=-\frac{16m}{\sqrt{1-v^2}}\left( 1-\omega _2\right) \eqno(2.17) $$
If the system consists of $n_1$ solitons and $n_2$ breather we have:
$$E=\sum\limits_{s=1}^{n_1}E_s+\sum\limits_{b=1}^{n_2}E_b,{~~} 
P=\sum\limits_{s=1}^{n_1}P_s+\sum\limits_{b=1}^{n_2}P_b,{~~} 
K=\sum\limits_{s=1}^{n_1}K_s+\sum\limits_{b=1}^{n_2}K_b. {~~}
\eqno(2.18)$$
Energy and momentum are related by following formula:
$$E^2=M^2+P^2.   \eqno(2.19) $$
Formulae (2.18) allow us to consider solitons and breathers as
elementary excitations of field $\varphi $.
\vskip 0.1 in
\begin{center}
{3. COLLISIONS OF LOCALIZED EXCITATIONS}
\end{center}
Due to the 1D of the problem the collision of solitons and breather take
place without dependence from values $x_{0s}$and $x_{0b}$ , $s=1,...,n_1,$ $
b=1,...,n_2$ because their velocities do not change. 
This peculiar property follows from energy and momentum conservation laws in
Sine-Gordon equation as well as other exactly integrable models (see [4,5]). 
Those conservation laws have the form:
$$v_1=v'_1;{~~~}v_2=v'_2. \eqno(3.1) $$
Here and later the values after collision will be devoted with a prime.

To analyze collision processes it is necessary to take into account the 
the "angular" momentum conservation law, which can be written as:
$$
x_1 E_1+x_2 E_2=x'_1 E'_1+x'_2 E'_2 .     \eqno(3.2)
$$ 
The result of collision is coordinate shifts (changes) and for breathers
phase shifts as well. Furthermore only pair collisions exist and there are
no many particles effects [4]. Therefore formulae (3.1),(3.2) provide the
general framework for studying the effects of collisions. In this section
conservation laws and coordinate $x_{01}$ and phase changes $\varphi _{01}$
for all type of collisions
will be written down explicitly from general formulas [4,5]. For simplicity
the index ''0'' will be omitted.

{\it Soliton-soliton collisions}. Conservation laws for two soliton
collision have the following form:
$$
v'_{1s}=v_{1s}, {~~}v'_{2s}=v_{2s}, {~~}
(x'_{1s}-x_{1s})E_{1s}+(x'_{2s}-x_{2s})E_{2s}=0 \eqno(3.3)$$
Coordinates shifts are given by:
$$\Delta x_{1s}=x'_{1s}-x_{1s}={\frac 4{E_{1s}}}sign(v_0)\ln |Z_{ss}|=
{\frac {\delta _{1s}} 2} sign(v_0)\ln |Z_{ss}|$$

$$\Delta x_{2s}=x'_{2s}-x_{2s}=-{\frac 4{E_{2s}}}sign(v_0)\ln |Z_{ss}|=
-{\frac {\delta _{2s}} 2}sign(v_0)\ln |Z_{ss}|,
\eqno(3.4)$$
where
$$Z_{ss}=\frac{1+\sqrt{1-v_0^2}}{1-\sqrt{1-v_0^2}}, {~~}
v_0=\frac{\left( v_{1s}-v_{2s}\right) }{\left( 1-v_{1s}v_{2s}\right) }. \eqno
(3.5)$$
Here $v_0$ is the velocity of relative motion of solitons which in the
general case can be written as:
$$v_0=\frac{\left( v_{1i}-v_{2k}\right) }{\left( 1-v_{1i}v_{2k}\right) },
{~~~~~}i,k=s,b,ph.\eqno(3.6)$$
{\it Soliton - breather collision.} In this case conservation laws can be
written as:
$$v_{1s}=v'_{1s},{~~~~}v_{2b}=v'_{2b},{~~~~}\omega
_{22}=\omega' _{22},$$
$$(x'_{1s}-x_{1s})E_{1s}+(x'_{2b}-x_{2b})E_{2b}=0. \eqno(3.7)$$
Coordinate shifts $\Delta x$ are described by the following formulae:
$$\Delta x_{1s}=x'_{1s}-x_{1s}={\frac 8{E_{1s}}}sign(v_0)\ln
|Z_{sb}| =\delta _{1s} sign(v_0) \ln |Z_{sb}|  $$
$$\Delta x_{2b}=x'_{2b}-x_{2b}=-{\frac 8{E_{2b}}}sign(v_0)\ln |Z_{ss}|=
-{\frac {\delta_{2b}} 2}sign(v_0)\ln |Z_{sb}|,     \eqno (3.8)$$
where:
$$
Z_{sb}=\frac{1+\omega _{2,b2}\sqrt{1-v_0^2}}{1-\omega _{2,b2}\sqrt{1-v_0^2}}
{~~}v_0=\frac{(v_{1s}-v_{2b})}{(1-v_{1s}v_{2b})}. \eqno(3.9)$$
The breather phase shift $(\Delta \varphi )_b$ is determined by the formula:
$$
\tan(\Delta \varphi )_b=-sign(v_0)\frac{\sqrt{1-v_0^2}\omega _{1,b2}}{v_0}.
\eqno(3.10)$$
Some comments regarding formulae (3.8) and (3.10) are in order. For
soliton coordinate shift (formula (3.8)) due to soliton - breather
collision one has a factor of ''8'' instead of ''4'' as in formula (3.4).
This difference can be explained as follows. If $\omega _{2,b2} \rightarrow
1, $ one can considered a breather as two solitons with opposite topological
charge. Since the coordinate shift does not depend on topological charge,
the soliton - breather collision in this case can be considered as a soliton
- two solitons collision, and thus the shift is doubled. Let us mention that
in the case $\omega _{2,b2}\rightarrow 1$, the quantities $Z_{ss}$ and $
Z_{sb}$ are the same.

Formula (3.10) implies that the phase shift is in general large: ($\Delta
\varphi )_b\sim \pi .$ It is small one only in the ultrarelativistic case $
|v_s-1|<<1.$ In this case ($\Delta \varphi )_{b1}\approx -sign(v_0)\sqrt{
1-v_0^2}\omega _{1,b1}\quad . $If $v_s<<1,$ then ($\Delta \varphi)_{b1}
\approx \pi /2.$

{\it Soliton-phonons collision}. This process is characterized by the
following conservation laws:
$$v'_{1s}=v_{1s},v'_{2ph}=v_{2ph},\omega' _{ph}=\omega_{ph},$$
$$(x'_{1s} - x_{1s}) E_{1s}+(x'_{2ph} - x_{2ph})E_{2ph}=0. \eqno(3.11) $$

The expressions for $x$ and $\varphi $ shifts are:
$$\Delta x_{2ph}=x'_{2ph}-x_{2ph}=-sign(v_0)\frac{m \sqrt{1-v_s^2}}
{\omega _{ph}(\omega _{ph}-kv_s)} \eqno(3.12) $$
$$\Delta x_{1s}=x'_{1s} - x_{1s}=-\frac{E_{2ph}}{E_{1s}} \Delta x_{2ph}
\eqno(3.13) $$
$$\tan (\Delta \varphi _{2ph})=-sign(v_0)\frac{\sqrt{1-v_0^2}}{v_0}. \eqno
(3.14) $$
Phonon energy $E_{2ph}$ and frequency $\omega _{ph}(k)$ are defined by the
formulae:
$$E_{2ph}=16\omega _2\omega _{ph}(k),\qquad \omega
_{ph}^2(k)=m^2+k^2\qquad. \eqno(3.15)$$
{\it Breather - breather collisions}. The corresponding conservation law
have the form:
$$
v'_{1b}=v_{1b},v'_{2b}=v_{2b},\omega' _{2,b1}=
\omega _{2,b1},\omega' _{2,b2}=\omega _{2,b2}$$
$$
(x'_{1b}-x_{1b})E_{1b}+(x'_{2b}-x_{2b})E_{2b}=0. \eqno(3.16)
$$
The expressions for the $x_b$ and $\varphi $ shifts are:
$$
\Delta x_{1b}=\frac 8{E_{1b}}sign(v_0)\ln |Z_{bb} Z_{bb}^{\prime }|=
{\frac {\delta_{1b}} 2} sign (v_0)\ln |Z_{bb} Z_{bb}^{\prime}|
$$
$$
\Delta x_{2b}=-\frac 8{E_{2b}}sign(v_0)\ln |Z_{bb}Z_{bb}^{\prime }|=
-{\frac {\delta_{2b}} 2} sign (v_0)\ln |Z_{bb} Z_{bb}^{\prime}|
\eqno(3.17)$$
$$
\tan (\Delta \varphi _{1b})=sign(v_0)\frac{2v_0\sqrt{1-v_0^2}\sin \psi
_1\cos \psi _2}{v_0^2+(1-v_0^2)(\cos ^2\psi _1-\cos ^2 \psi _2)}.\eqno(3.18)$$
$$
\tan(\Delta \varphi _{2b})=-sign(v_0)\frac{2v_0\sqrt{1-v_0^2}\sin \psi
_2\cos \psi _1}{v_0^2+(1-v_0^2)(\cos ^2\psi _1-\cos ^2\psi _2)}.\eqno(3.19)$$
Here:
$$
Z_{bb}=\frac{1{-}\sqrt{1-v_0^2}\cos(\psi _1+\psi _2)}{1{-}\sqrt{1-v_0^2}
\cos(\psi _{1} - \psi _2)}, \quad  
Z_{bb}^{\prime }=\frac{1{+}\sqrt{1-v_0^2}\cos (\psi _1-\psi _2)}{1 +
\sqrt{1-v_0^2}\cos (\psi _1+\psi _2)}.\eqno(3.20)$$

The angles $\psi _1$ and $\psi _2$ are related to the parameters $\omega
_{2,b1}$ and $\omega _{2,b2}$ by the formulae:
$$\sin \psi _1=\omega _{2,b1},{~~~} \sin \psi _2=\omega _{2,b2} \eqno(3.21)$$
{\it Breather - phonon collision}. Since phonons are the limiting case of
breather when $\omega _2\rightarrow 0$, formulae for breather - phonons
collision can be obtained from ones in the previous paragraph by passing to
the limit $\omega _{2,b2}\rightarrow 0$. It means that the second breather
reduces to phonon wave packet and the notion of group velocity and
coordinate of wave packet automatically appears. The breather velocity $v_b$
reduces to the group velocity of the wave packet and the coordinate $x_b$
reduces to the coordinate of the center of the wave packet.

The corresponding conservation law has the form:
$$
v'_{1b}=v_{1b},{~~~}v'_{2ph}=v_{2ph},{~~~}\omega'_{2,b1}=
\omega _{2,b1}, \omega' _{2,ph2}=\omega _{2,ph2}$$
$$(x_{1b}^{\prime }-x_{1b})E_{1b}+(x_{2ph}^{\prime }-x_{2ph})E_{2ph}=0,
\eqno(3.22)$$
where:
$$\Delta x_{2ph}=-sign(v_0)\frac{2 m\omega_{2, b1}\sqrt{1-v_b^2}}{\omega
_{ph} [\omega _{ph}-kv_b][1-(1-v_0^2)\omega _{1,b1}^2]},\qquad 
\Delta x_{1b}=-\frac{E_{2ph}}{E_{1b}} \Delta x_{2ph}  \eqno(3.23)$$

$$\tan (\Delta \varphi _{1b})=-sign(v_0)\frac{2v_0\sqrt{1-v_0^2}\omega _{2,b1}}
{2v_0^2-1+(1-v_0^2)\omega _{1,b1}^2}\eqno(3.24)$$
$$\Delta \varphi _{2ph}=-sign(v_0)\frac{2v_0\sqrt{1-v_0^2}\omega
_{2,ph2}\omega _{1,b1}}{2v_0^2-1+(1-v_0^2)\omega _{1,b1}^2}. \eqno(3.25)$$

{\it Phonon-phonon collisions}. As it is well known, in linear theory
phonon-phonon collisions are absent. Here phonon - phonon interaction is the
result of nonlinearity of the Sine-Gordon equation. Corresponding relations
can be obtained from ones in previous paragraph by passing the limit $\omega
_{2,b1}\rightarrow 0.$ By considering $\omega _{2,b1}$ as a small quantity, $
\omega _{2,b1}<<1,$ one has:
$$
v_{1ph}^{\prime }=v_{1ph},{~~~~}v_{2ph}^{\prime }=v_{2ph}, 
{~~~}\Delta x_{1ph}E_{1ph}+\Delta x_{2ph}E_{2ph} =0. \eqno(3.26)$$
The shift of phonons coordinate is:
$$
\Delta x_{1ph}=sign(v_0)\frac{32}{E_{1,ph}}\frac{\sqrt{1-v_0^2}\omega
_{2,ph2}\omega _{2,,ph1}}{v_0^2}\qquad \Delta x_{2ph}=-\frac{E_{1ph}}{E_{2ph}}
\Delta x_{1ph}, \eqno(3.27)$$
where:
$$
E_{1ph}=16\omega _{21ph}\omega _{1ph},{~~~~}E_{2ph}=16\omega _{22ph}\omega
_{2ph}.\eqno(3.28)$$
The change of phonon phase is:
$$\Delta \varphi _{1ph}=-2sign(v_0)\omega _{2,ph1}\frac{\sqrt{1-v_0^2}}{v_0}
,\qquad \Delta \varphi _{2ph}= - 2sign(v_0)\omega _{2,ph2}\frac{\sqrt{1-v_0^2}
}{v_0}.                \eqno(3.29)$$

The examination of phonons as a limiting case of breathers gives the
possibility to consider changes of phonon phase and coordinate simultaneously.

Let us analyze phase shifts of breathers and phonons. It is easy to see from
formulae (3.10)-(3.12) and (3.14) that breather and phonon shifts are small
for soliton-breather and soliton-phonon collisions in the general case. For
breather - phonon collisions breather and phonon shifts are essentially
different (see (3.25),(3.29)). Breather phase shift are proportional to $
\omega _{2ph}$ and therefore small. Phonon phase shift ($\Delta \varphi
)_{ph}\approx 1$. For phonon - phonon collisions phase shifts are small (see
(3.33)). This agrees with the standard linear theory of phonons. In
accordance with the usual theory of phonons shifts of phonon wave packets do
not take place.

In conclusion, let us emphasize that soliton topological charge is conserved
for each type of collisions. The processes leading to soliton-antisoliton
bound state formation and the reverse processes do not take place because of
energy conservation. 
Thus in the framework of the exactly integrable Sine-Gordon model with
Hamiltonian (2.3) the numbers of solitons, antisolitons, breathers and
phonons are defined by initial conditions at each point of $(x,t)$.
Interactions destroying the integrability of the system lead to possibility
of energy exchange between the LE and to the formation and decay
of breathers. In the present paper only the processes with conservation of
energy and topological charge have been considered. The processes with
energy exchange will be analyzed in a future publication.
\vskip 0.1 in
\begin{center}
{4. PROBABILITY OF SCATTERING PROCESSES}
\end{center}
\vskip 0.1 in
To construct collision integrals for processes considered in the present
section let us analyze the probability of the corresponding scattering
process. The simplest process is soliton - soliton scattering. By definition
the initial state {\it (i)} of a soliton pair is described by its velocities
and center of gravity coordinates:  
$$i\equiv \left(
x_1,v_1,x_2,v_2\right) \equiv \left( 1,2\right) .
\eqno(4.1)$$
The final state {\it (f)} is defined by:  
$$f\equiv (x'_1,v'_1 ,x'_2 ,v'_2 )\equiv 
\left( 1' ,2' \right). \eqno(4.2)$$
Current density of solitons per unit density is given by: 
$$j_1=\left| v_0\right|. \eqno(4.3)$$
Taking into account formula (4.3) and the fact that two solitons collide in
any case, the transition probability
per unit time from state $(i)$ to state $(f)$ can be presented as:
$$W_{i\rightarrow f}\equiv W(1^{\prime },2^{\prime }|1,2)=|v_0|\delta
(v_1^{\prime }-v_1)\delta (v_2^{\prime }-v_2)\delta ({\frac{E_1}{E_2}}
(x_1^{\prime }-x_1)+$$
$$
+(x_2^{\prime }-x_2))\delta (x_1^{\prime }-x_1-\Delta x(v_1,v_2)) .
\eqno(4.4)
$$
In this formula the first two delta functions describe energy conservation
laws, the third one - ''angular'' momentum conservation law, and last delta
function describes coordinate shift. The coordinate shift of second soliton
follows from the ''angular'' momentum conservation law. The coordinate shift 
$\Delta $  $x_1$ is defined by expression (3.4),(3.5).

It is convenient to rewrite formula (4.4) in the form:

$$W(1^{\prime },2^{\prime }|1,2)=$$

$$=R_{ss}(v_1^{\prime },v_2^{\prime }|v_1,v_2)\delta (x_1^{\prime
}-x_1-\Delta x_1(v_1,v_2))\delta (x_2^{\prime }-x_2-\Delta x_2(v_1,v_2)) .
\eqno(4.5)$$
where: 
$$ R_{ss} (v_1^{\prime },v_2^{\prime }|v_1,v_2)=|v_0|\delta (v_1^{\prime}
-v_1)\delta (v_2^{\prime }-v_2).    \eqno(4.6)$$
Let us note that: 
$$R_{ss}\left( v_1^{\prime },v_2^{\prime }|v_1,v_2\right) =
R_{ss}\left( v_1,v_2|v_1^{\prime },v_2^{\prime }\right)
=R_{ss}(v_2^{\prime },v_1^{\prime }|v_2,v_1). \eqno(4.7)$$
For the probability of the inverse process the following expression can be
written:
$$W(1,2|1^{\prime},2^{\prime})=$$
$$=R_{ss}(v_1^{\prime },v_2^{\prime }|v_1,v_2)\delta (x_1-x_1^{\prime}
+\Delta x_1(v_1,v_2))\delta (x_2-x_2^{\prime }+\Delta x_2(v_1,v_2)) .
\eqno(4.8)$$
Let us emphasize that the probability defined by formula (4.4) does not
satisfy the detailed balance principle in the  standard form: 
$$\tilde{W}(1',2'|1,2)= \tilde{W}(1^{*},2^{*}|{1'}^{*},{2'}^{*}) .
\eqno(4.9)$$

In (4.9) the following notation has been used: $(1 ^{*} )\equiv ( x_{1} ,-
v_{1}) $ .



For further analysis it is convenient to present the expression (4.7) for
the probability $W$ as a sum of two parts $W^r$ and $W^m$:

$$W_{ss}(1,2|1',2')=W_{ss}^r(1,2|1',2')+W_{ss}^m(1,2|1',2'), 
\eqno(4.10)$$
where:
$$W_{ss}^r={\frac{1 }{2}} R_{ss}[\delta(x_{1s}-x^{\prime}_{1s}-\Delta x_{1s})
\delta(x_{2s}-x^{\prime}_{2s}-\Delta x_{2s})-$$

$$-\delta(x_{1s}-x^{\prime}_{1s}+\Delta x_{1s})
\delta(x_{2s}-x^{\prime}_{2s}+\Delta x_{2s})] \eqno(4.11)$$

$$W_{ss}^m={\frac 12}R_{ss}[\delta (x_{1s}-x_{1s}^{\prime }-\Delta
x_{1s})\delta (x_{2s}-x_{2s}^{\prime }-\Delta x_{2s})+$$
$$+\delta(x_{1s}-x^{\prime}_{1s}+\Delta x_{1s})
\delta(x_{2s}-x^{\prime}_{2s}+\Delta x_{2s})]. \eqno(4.12)$$
The probability $W_{ss}^r$, as it will be shown later, describes the effect
of {\it renormalization} of soliton velocity, and the probability $W_{ss}^m$
describes the homogenization of solitons phenomena i.e. the effect of mixing 
of states. The quantity $W_{ss}^m$ is related with entropy production in 
the soliton gas and the kinetic coefficients (see sections 5,6).

It is easy to convince oneself that: 
$$W_{ss}^m(1^{\prime },2^{\prime }|1,2)=W_{ss}^m(1,2|1^{\prime },2^{\prime }). 
\eqno(4.13)$$
This equality means that for the {\it dissipative} part of the transition
probability the usual detailed balance principle is valid.

Let us consider now the general case of LE $i$ and $k$, 
($i,k=s,b,ph)$ for soliton, breather and phonon correspondingly. The
probability of such type of collisions per unit time $W_{ik}$ can be defined
by following formula: 
$$W_{ik} \equiv W_{ik}(1^{\prime},2^{\prime}|1,2) =$$
$$=R_{ik} \delta ( X^{\prime}_{i}-X_{i}-d_{i}( 1,2))
\delta(X^{\prime}_{k}-X_{k}-d_{k}(1,2)) ,
\eqno(4.14)$$
where:
$$R_{ik}=R_{ik}(V_i^{\prime },V_k^{\prime }|V_i,V_k)=|v_0|\delta
(V_i-V_i^{\prime })\delta (V_k-V_k^{\prime }) .
\eqno(4.15)$$
In formulae (4.14),(4.15) the following notations have been used. Numbers 1
and 2 mean the set of variables defining of the colliding LE
states, $d_i$, is the shift of LE coordinate or phase. In the
case of solitons $1\equiv ( x_{s} ,v_{s} ) $, for breather or phonon $1\equiv (
x_{i} ,\varphi_{i} ,v_{i} ,\omega_{2 i}) ,$ $i= b, ph$. For simplicity two
component coordinate $X_i$ and two component velocity $V_i$ have been
introduced in the following way: 
$$X_s=x_s,V_s=v_s,\qquad X_i=(x_i,\varphi _i),V_i=(v_i,\omega _{2i}),i=b,ph.
\eqno(4.16)$$
In (4.15), the delta function, $\delta $ ( $V_{i} -V^{\prime}_{i} )$, describes
conservation law of energy and momentum of each LE; in the expression
(4.14), the delta function $\delta $ $(X^{\prime}_{i} -X_{i} -d_{i} (1,2))$
defines the coordinate and phase for breather and phonon shifts after
collision.

>From the definition of $R_{ik}$ it follows that :
$$R_{ik} =R( V^{\prime}_{i} ,V^{\prime}_{k}|V_{i} ,V_{k})
=R(V_{i},V_{k}|V^{\prime}_{i},V^{\prime}_{k}). \eqno(4.17)$$
Later only the case of weak inhomogeneous gas of LE will be
considered, i.e. when the characteristic length $l$ of the distribution 
function is much larger than the shift $\Delta X_{i} $: 
$$|l_i|>>|\Delta X_i| \eqno(4.18)$$

The mean free path of LE is:
$$ l \sim  {1 \over n},         \eqno(4.20) $$
because the cross section of scattering is equal to $1$.

The coordinate shifts of LE are of the order of $\delta_i$
(see section 3), where
$$ \delta _i (v,\omega_2) \sim {1 \over {m(v) \omega_2}} \eqno(4.22) $$
Therefore the condition (4.18) rewritten in the form:

$$ {1 \over n_i}>>\delta _i{~~~}or{~~~}m(v) \omega_2 >> {1 \over l} \sim  n_i
{~~~}i=s,b,ph ,
\eqno (4.21) $$

means that average distance between LE much larger their sizes.
In other words the condition of small inhomogeneous (4.18) coincides with the
condition of small density of LE gas.

Let us present $W_{ik}$ as a sum of two parts $W_{ik}^r$ and $W_{ik}^m$
by the same way as (4.10)-(4.12). Assuming that (4.18) is valid, we 
can expand the delta functions in the expression (4.14) for $W_{ik}(1,2|1,2)$
in powers series $\Delta X_i$. Then the expressions for $W_{ik}^r$ and
$W _{ik}^m$ become:
$$W_{ik}^r(1^{\prime },2^{\prime }|1,2)=R_{ik}(1^{\prime }2^{\prime },
|1,2)[d_i{\frac \partial {\partial X_i}}+d_k{\frac \partial {\partial X_k}}]
\delta(X_i-X_i^{\prime })\delta (X_k-X_k^{\prime })\eqno(4.22) $$
$$W_{ik}^m(1^{\prime },2^{\prime }|1,2)= $$
$$=R_{ik}(1^{\prime }2^{\prime },|1,2)[2+d_i^2{\frac{\partial ^2}{\partial
X_i^2}}+d_k^2{\frac{\partial ^2}{\partial X_k^2}}+2d_id_k{\frac{\partial ^2}{
\partial X_i\partial X_k}}]\delta (X_i-X_i^{\prime })\delta (X_k-X_k^{\prime
}). \eqno(4.23)$$
\vskip 0.1 in
\begin{center}
{5.COLLISION INTEGRALS}
\end{center}
In the present section the expressions for the collision integrals for all type
of LE  in SG equation will be derived. It is easier to do
this starting from the formulae for probabilities of collisions. Let us
introduce following notations:
$$f=f(X,V,t),B=B(X,V,t),N=N(X,V,t). \eqno(5.1)$$
for the distribution functions of solitons, breathers and phonons as {\it f, B}
and {\it N} correspondingly. The probability to find the LE in
the state $(1,1+d1)$ can be defined the usual way:

$$dW_i=F_i(1)d1,\eqno(5.2)$$
where\\
$i=s,b,ph$ and $F_s=f,F_b=B,F_{ph}=N$.
The total result of collisions of a sample LE with the other
LE is the sum of each collision. This is due to the special type
of interaction in exactly integrable models.
Therefore the general collision integral can be presented as a sum of
partial collision integrals. For example in the case of solitons the
collision integral can be written as:
$${\cal L}_s={\cal L}_{ss}\{f,f\}+{\cal L}_{sb}\{f,B\}+{\cal L}_{sph}\{f,N\},
\eqno (5.3)$$
where ${\cal L}_{ss},{\cal L}_{sb},{\cal L}_{sph}$ are soliton-soliton,
soliton-breather and soliton-phonon collision integrals respectively, which
have the following form:

$${\cal L}_{ss}\{f,f\}=\int d1^{\prime }d2^{\prime }d2\{W_{ss}(1,2|1^{\prime
},2^{\prime })f(1^{\prime })f(2^{\prime })-W_{ss}(1^{\prime },2^{\prime
}|1,2)f(1)f(2)\} \eqno (5.4)$$

$${\cal L}_{sb}\{f,B\}=\int d1^{\prime }d2^{\prime }d2\{W_{sb}(1,2|1^{\prime
},2^{\prime })f(1^{\prime })B(2^{\prime })-$$

$$-W_{sb}(1^{\prime },2^{\prime }|1,2)f(1)B(2)\} \eqno (5.5)$$

$${\cal L}_{sph}\{f,N\}=\int d1^{\prime }d2^{\prime
}d2\{W_{sph}(1,2|1^{\prime },2^{\prime })f(1^{\prime })N(2^{\prime })-{~}$$
$$-W_{ss}(1^{\prime },2^{\prime }|1,2)f(1)N(2)\}.\eqno(5.6)$$
The first term in formula (5.4) and in formulae (5.5) and (5.6) describes
solitons  ''arriving'' at the state $(1)$ as the
result of collisions and second term describes solitons ''leaving'' this state.

The general expression for the collision integral has the form:

$${\cal L}_i=\sum_k{\cal L}_{ik}\{F_iF_k\} .\eqno(5.7)$$

Here:

$${\cal L}_{ik}\{F_iF_k\}=\int d1^{\prime }d2^{\prime
}d2\{W_{ik}(1,2|1^{\prime },2^{\prime })F_i(1^{\prime })F_k(2^{\prime })-$$
-$$W_{ik}(1^{\prime },2^{\prime }|1,2)F_i(1)F_k(2)\} .\eqno(5.8)$$
Let us discuss ones more the detailed balance principle for usual particles 
in the form:

$$\tilde{W}(1^{\prime },2^{\prime }|1,2)=
\tilde{W}(1,2|1^{\prime },2^{\prime }) \eqno(5.9)$$

In this form detailed balance principle describes two particle collision
(generalization to three, four, ... particle collision is well known).
Formula (5.9) means that ''arriving'' number in state $(1^{\prime
},2^{\prime })$ from state $(1,2)$ is equal to the ''leaving'' number from
state $(1^{\prime },2^{\prime })$ to state $(1,2)$. If the total number of
''arriving'' particles to some fixed state $(1,2)$ is equal to the total
number of ''leaving'' particles from state $(1,2)$ then new {\it 
smoothened local balance principle} can be formulated as:

$$\int W_{ik}(1,2|1^{\prime },2^{\prime })d1^{\prime }d2^{\prime }=\int
W_{ik}(1^{\prime },2^{\prime }|1,2)d1^{\prime }d2^{\prime } \eqno (5.10)$$
where $i,k=s,b,ph.$

It is not difficult to show that the probabilities of collisions defined in
section 3 satisfy this condition. Thus the probabilities, $W_{ik}$, of
LE scattering processes in SG model satisfy the smoothened local balance
principle
smoothened local balance principle (5.10), while the dissipation parts of
probabilities $W_{ik}^m$ satisfy detailed balance principle (5.9).

The distribution functions $f,B,N$ in thermodynamic equilibrium must satisfy
the following condition:
$${\cal L}_{ik}\{F_i,F_k\}=0,\eqno(5.11)$$
in accordance with the smoothened local equilibrium principle for each $i-k$
collision integral. It is necessary to emphasize that smoothened local balance
principle (5.10) puts limitation on the probabilities of collisions, on the 
thermodynamic equilibrium condition (5.11) and on the distribution functions 
$F_i$.

Having mentioned the general properties of collision integrals of
LE in the SG equation, we proceed with the examination of each
one. From formulas (4.21)- (4.22) it is easy to obtain the following expression
for the collision integral for solitons in the low density case :

$${\cal L}_s\{f_1,F_{2i}\}=-\delta v_s{\frac{\partial f_1}{\partial x}}+{\cal D}
_s(v_s){\frac{\partial ^2f_1}{\partial x^2}}, \eqno(5.12)$$

where renormalization of soliton velocity $\delta v$ and local coefficient
of self-diffusion ${\cal D}_s(v)$ are given by summing the partial
contributions: 
$$\delta v_s=\sum _i\delta v_{si}{~~~}{\cal D}_s(v_s)=\sum _i{\cal D}
_{si}(v_s). \eqno (5.13) $$
Here 
$$\delta v_{si}=\int |v_0|\Delta x_{si}(v_{1s},v_{2i})F_id2 \eqno(5.14)$$

$${\cal D}_{si}={\frac 12}\int |v_0{}|[\Delta x_{si}(v_{1s},v_{2i})]^2F_id2
\eqno(5.15)$$

The collision integrals for breathers can be analyzed by a similar way, but
with a very important difference, connected with the conditions (4.20) and
(4.21). The simplest case of breather ensemble is a ensemble with
distribution function of the form: 
$$B(x,\varphi ,v,\omega _2,t)=B(x,\varphi,v,t)\delta (\omega _0-\omega _2),
\eqno(5.16)$$
where $\omega _0$ and density of particles $n_i$ satisfy the condition: $
\omega _0m(v)>>n_i.$

It is possible to consider more general breather distribution functions, e.g.
$$B=B(x,\varphi ,v,t)b(\omega ),$$
where $b(\omega )$ has a sharp maximum near the point $\omega =\omega _0$.
For simplicity only the case (5.16) will be considered. In this case all
integrations under $\omega _2$ are trivial.

For breathers the collision integral has the following form:
$${\cal L}_b\{B_1,F_{2i}\}=-\delta v_b{\frac{\partial B_1}{\partial x}}
-\delta \omega _b{\frac{\partial B_1}{\partial \varphi }}+{\cal D}_b{\frac{
\partial ^2B_1}{\partial x^2}}+{\cal F}_b{\frac{\partial ^2B_1}{\partial
\varphi ^2}}+2{\cal K}_b{\frac{\partial ^2B_1}{\partial x \partial \varphi }}.
\eqno (5.17)$$

The first two terms in this formula describe renormalization of breather
velocity $\delta v_b$ and its internal oscillation frequency $\delta \omega
_b$. The last three terms describe self-diffusion in $(x,\varphi )$ space.
As in the soliton case, the quantities $\delta v_b,\delta \omega _b,{\cal D}_b,
{\cal F}_b$ and ${\cal K}_b$ are sums of partial contributions:

$$\delta v_b=\sum _i(\delta v_b)_i,\delta \omega _b=\sum _i(\omega _b)_i,
{\cal D}_b=\sum _i({\cal D}_b)_i,{\cal F}_b=\sum _i({\cal F}_b)_i,{\cal K
}_b=\sum _i({\cal K}_b)_i ,\eqno (5.18)$$

where:

$$\delta v_{bi}=\int |v_0|\Delta x_{bi}(v_{1b},v_{2i})F_{2i}d2 \eqno(5.19)$$
$$\delta \omega _{bi}=\int |v_0|\Delta \varphi _{bs}(v_{1b},v_{2i})F_{2i}d2
\eqno(5.20)$$
$${\cal D}_{bi}={\frac 12}\int |v_0|[\Delta x_{bi}(v_{1b},v_{2i})]^2F_{2i}d2
\eqno(5.21)$$
$${\cal F}_{bi}={\frac 12}\int |v_0|[\Delta \varphi
_{bi}(v_{1b},v_{2i})]^2F_{2i}d2 \eqno(5.22)$$
$${\cal K}_{bi}={\frac 12}\int |v_0|\Delta x_{bi}(v_{1b},v_{2i})\Delta
\varphi _{bi}(v_{1b},v_{2i})F_{2i}d2 .\eqno(5.23)$$
\vskip 0.1 in
\begin{center}
{6.KINETIC EQUATIONS AND ENTROPY PRODUCTION}
\end{center}

The Boltzmann type kinetic equations for LE with
collision integrals constructed in previous subsection can be written
following the standard way as:

$${\frac{\partial f}{\partial t}}+(v+\delta v_s(v)){\frac{\partial f}
{\partial x}}
={\cal D}_s(v){\frac{\partial ^2f}{\partial x^2}} \eqno (6.1)$$

$${\frac{\partial B}{\partial t}}+(v+\delta v_b(V)){\frac{\partial B}
{\partial x}}+(\omega +\delta \omega _b(V)){\frac{\partial B}{\partial 
\varphi }}={\cal D}_b(V){\frac{\partial ^2B}{\partial x^2}}+2{\cal K}_b(V)
{\frac{\partial ^2B}{\partial x\partial \varphi }}+{\cal F}_b(V)
{\frac{\partial ^2B}{\partial\varphi ^2}} \eqno (6.2)$$

$${\frac{\partial N}{\partial t}}+(v+\delta v_{ph}(V)){\frac{\partial N}{
\partial x}}+(\omega+\delta \omega_{ph}(V)){\frac{\partial N}{\partial
\varphi }}={\cal D}_{ph}(V){\frac{\partial ^2N}{\partial x^2}}+2{\cal K}_{ph}
(V){\frac{\partial ^2N}{\partial x\partial \varphi }}+{\cal F}_{ph}(V){\frac{
\partial ^2N}{\partial \varphi ^2}} \eqno (6.3)$$

Here terms from collision integrals describing velocity renormalization have
been written on the left of eqs. (6.1)-(6.3). On the right of these
equations there are only those terms describing dissipative processes leading to
homogenization of distribution functions of LE.

It is necessary to emphasize that collision integrals in kinetic equations
(6.1) - (6.3) are equal to zero in the homogeneous case. Therefore the 
stationary solution of kinetic equations have following form:
$$f=f(v),{~~~~}B=B(v,\omega _2). \eqno (6.4)$$
Here $f(v),B(v,\omega _2)$ are arbitrary functions of its arguments.

In other words the kinetic equations (6.1)-(6.2) describe {\it homogenization
} i.e. mixing of distribution function of LE up to homogeneous
state in real space (for breathers - in $(x,\varphi )$ space) and
demonstrate that chaotization in momentum space cannot be realized.

For chaotization processes in momentum space it is necessary to
exceed the limits of exactly integrable model i.e. to take into account the
terms destroying the integrability in the Hamiltonian of the system.

Let us show now that homogenization of distribution function leads to
entropy production in the SG localized excitations gas. The entropy of
classical soliton gas and boson gases of breathers and phonons can be
defined by standard way:
$$S=\sum _k S_k,{~~~}k=s,b,ph \eqno (6.5)$$
$$S_k=-\int F_kln(F_k/e)d1 .\eqno (6.6)$$
The entropy evolution in time is described by the formulae:

$${\frac{dS_k}{dt}}=-\int \frac{\partial F_k(1)}{\partial t}\ln F_k(1)d1.
\eqno (6.7)$$

Using kinetic equations (6.1)-(6.3), definitions of ${\cal D,K}$ and 
${\cal F}$ one can find that:
$${\frac{dS_k}{dt}}=\int q_kd1,\eqno(6.8)$$

where the source q of the entropy production is:
$$q_k=\int \sum_i|v_0(1,2)|_{ki}\frac{F_i}{F_k}\{[\Delta x(1,2)]_{ki}\frac{
\partial F_k(1)}{\partial x}+[\Delta \varphi (1,2)]_{ki}\frac{\partial F_k(1)}{
\partial \varphi }\}^2d2 . \eqno (6.9)$$

It is obviously that the expression is positive. It means that (6.9)
proves the Boltzmann entropy production theorem. We would like to emphasize
that the entropy production is connected only with inhomogeneity in real
space (x,$\varphi ).$It is easy to see that in homogeneous case $\partial
F/\partial x = \partial F/\partial \varphi =0$ and there no entropy production.

\vskip 0.1 in
\begin{center}
{7.TRANSPORT EQUATIONS}
\end{center}
In the present subsection the consequences from kinetic equations
(6.1)-(6.3) will be analyzed. Let us emphasize that the homogenization of
LE in real space means the homogenization of LE in
temperature, concentration and macroscopic velocity spaces. The local
macroscopic temperature $T(x)$ is defined trough the local energy $E(x)$
averaged over a distance $d_0$ around $x$ satisfying the inequality $|\Delta
X_i| << d_0$. In other words collisions of LE lead to creation
of diffusion, thermoconductivity and intrinsic friction processes. It is
easy to note that the transport equations will have the form of local
conservation laws for each type of LE separately. In Sine-Gordon
system the numbers of solitons, breathers and phonons are conserved
separately. Besides, momentum, energy and angular velocity $\frac{\partial
\varphi }{\partial t }$ (in the breather case) of each LE are
conserved in each collision. The transport equation can be written in the
following general form.

For the {\it solitons}:

$${\frac \partial {\partial t}}n_s<a_s>+{\frac \partial {\partial x}}
[U_s^r+U_s^m]=0. \eqno (7.1)$$

For the {\it breathers}:

$${\frac \partial {\partial t}}n_b<a_b>+{\frac \partial {\partial x}}
[U_b^r+U_b^m]+{\frac \partial {\partial \varphi }}[W_b^r+W_b^m]=0.\eqno(7.2)$$

In formulae (7.1)-(7.2) following notation have been used.

For {\it solitons}:

$$n_s<a_s>=\int a(x,v)f(x,v,t)dv \eqno (7.3)$$

$$U_s^r=\int a(x,v)[v+\delta v]f(x,v,t)dv \eqno(7.4)$$

$$U_s^m=-{\frac \partial {\partial x}}\int a(x,v)
{\cal D}_sf(x,v,t)dv. \eqno (7.5)$$

For {\it breathers}:

$$n_b<a_b>=\int a(x,v,\varphi ,\omega )B(x,v,\varphi ,\omega ,t)dvd\omega
\eqno (7.6)$$

$$ U_b^r=\int a(x,v,\varphi ,\omega )[v+\delta v]
B(x,v,\varphi ,\omega ,t)dv d\omega \eqno (7.7)$$

$$U_b^m=-{\frac \partial {\partial x}}\int a(x,v,\varphi ,\omega)
{\cal D}_bB(x,v,\varphi ,\omega ,t) dv d\omega - $$
$$ -{\frac \partial {\partial \varphi }}\int a(x,v,\varphi ,\omega)
{\cal K}_bB(x,v,\varphi ,\omega ,t)dv d\omega \eqno (7.8)$$

$$W_b^r=\int a(x,v,\varphi ,\omega )[\omega+\delta \omega]
B(x,v,\varphi ,\omega ,t)dvd\omega 
                                                            \eqno (7.9)$$

$$W_b^m=-{\frac \partial {\partial \varphi }}\int a(x,v,\varphi ,\omega)
{\cal F}_bB(x,v,\varphi ,\omega ,t)dvd\omega -$$

$$-{\frac \partial {\partial x}}\int a(x,v,\varphi ,\omega )
{\cal K}_bB(x,v,\varphi,\omega ,t)dvd\omega. \eqno (7.10)$$

Substituting $1$ , velocity and energy of LE for the variable $a $
it is easy to obtain following transport equations: continuity equations,
hydrodynamics equations and equations for local energy density. Let us write
these equations in the explicit form.

{\it Continuity equations} ($a=1$) can be written as:

$${\frac{\partial n_s}{\partial t}}+{\frac \partial {\partial x}}
(j_s^r+j_s^m)=0\eqno (7.11)$$

$${\frac{\partial n_b}{\partial t}}+{\frac \partial {\partial x}}
(j_b^r+j_b^m)+{\frac \partial {\partial \varphi }}(i_b^r+i_b^m)=0. \eqno
(7.12)$$

Here the standard notation $j_i$ and $i_i$ have been used for $U_i$ and $W_i$
with $a=1$ respectively.

{\it Hydrodynamics equations} can be derived from equations (7.1),(7.2)
with $a=v$ and $(v,\omega _{2b})$ for solitons and breather respectively
and have the following form:

$${\frac \partial {\partial t}}n_su_s+{\frac \partial {\partial x}}
(P_s^r+P_s^m)=0,\eqno (7.13)$$
$$
{\frac{\partial }{\partial t}} n_b u_b + {\frac{\partial }{\partial x}} (P_b
^{ r}+P_b^{ m})+{\frac{\partial }{\partial \varphi}} (\Pi_b^{ r} +\Pi_ b^{
m})=0 \eqno (7.14) $$
$$
{\frac{\partial }{\partial t}} n_b w_b + {\frac{\partial }{\partial x}} (Q_b
^{ r}+Q_b^{ m})+{\frac{\partial }{\partial \varphi}} (R_b^{ r} +R_ b^{ m})=0.
\eqno (7.15) $$

Here $u_s$ and $u_b$ are hydrodynamic velocities of solitons and breathers
correspondingly, $w$ is hydrodynamic velocity in $\varphi $ space, $P_i^r$
and $P_i^m,i=s,b$ are pressures for solitons and breathers, the 
quantities $\Pi _b^r$ and $\Pi _m^d$ 
are pressures of breathers due to inhomogeneities
in $\varphi $ space, values $Q$ and $R$ are defined by formulas (7.7)-(7.10)
with $a=\omega _{2b}$. 

{\it The energy transport equations} can be derived when $a=E_i$ for
solitons and breathers.
These equations can be written as: 

$${\frac \partial {\partial t}}n_sT_s+{\frac \partial {\partial x}}
(U_s^r+U_s^m)=0,     \eqno(7.16) $$
$$
{\frac{\partial }{\partial t}} n_b T_b + {\frac{\partial }{\partial x}} (U_b
^{ r}+U_b^{m})+{\frac{\partial }{\partial \varphi}} (W_b^{r} +W_ b^{m})=0 .
\eqno (7.17) $$

The quantities $T_i$ means the average energies of corresponding LE.
When $u_i=w_i=0$ the quantities $T_i$ are the average energies of chaotic 
motion. $U_i$ and $W_i$ are energy density currents; one can conclude 
that only $U_i^m $ and $W^m$ are different from 0 when $u_i=w_i=0$.

Let us emphasize the important property of transport equations
(7.11)-(7.17). It is easy to see that for any homogeneous distribution
functions: 
$$f = f(v,t), {~~~}B = B(v, \omega_{2b}, t) \eqno (7.18) $$
with constant temperatures, hydrodynamic velocities and chemical potentials $
\mu_i$ then all dissipative terms in these equations are equal to zero. In
other words there no energy and momentum exchange between homogeneous gases
of solitons and breathers.

This special property of equations (7.14) - (7.16) is eliminated by taking
into consideration terms in Hamiltonian which destroy the integrability of
the model.

As an example of explicit calculation of transport coefficients we shall 
obtain the expression for the self-diffusion coefficients assuming that the
distribution function of solitons and breathers have following form: 

$$f=C_se^{-{\frac{m_sv_s^2}{2k_BT}}}, {~~~~}
B=C_be^{-{\frac{E_b}{k_BT}}}\delta (\omega _2-\omega _0). \eqno(7.19) $$

Let us discuss the concrete expressions for solitons and breathers diffusion
coefficients due to soliton-solitons, breather - breathers and solitons
breathers collisions. For simplicity we will consider the nonrelativistic
solitons and breathers only ($v<<1).$ This case corresponds to small
temperatures $T<<m.$

The diffusion current of solitons can be presented as:

$$j_s^d=\frac{\partial n_s}{\partial x}[n_s D_{ss}+n_b D_{sb}]+
\frac{\partial n_b}{\partial x}n_s D_{sb}  \eqno(7.20)$$

Using formulae for $(\Delta \varphi )_{ik}$ and $(\Delta x)_{ik}$ from
paragraph 3 , formulae (5.13), (5.18), (7.5), and (7.19) it is possible to
calculate both $ D_{ss}$ and $ D_{sb}.$ We will present here
the final results:

$$ D_{ss}=(1/4) \delta_s^2 (T/\pi M_s)^{1/2}\{[\ln
(\gamma M_s/T)]^2 +C \}                            \eqno(7.21)$$
$$ D_{sb} = (1/2) \delta_s^2 (2T / \pi \mu _{sb})_{sb}^{1/2}I_{sb} ,
                                                   \eqno(7.22) $$
where:

$$I_{sb} = (\ln \frac{1+\omega _0}{1-\omega _0})^2,\quad \quad if\quad 
T/\mu_{sb} \ll 1-\omega _0^2 $$
$$I_{sb}=[\ln (2 \gamma \mu _{sb}/T)]^2+C 
\quad \quad if\quad \quad 1\gg T/\mu _{sb}\gg 1-\omega _0^2. \eqno(7.23)$$
Here:

$$C\approx 1.6,{~}\gamma \approx 1.8.$$

There are two dissipative currents for breathers $j_b^d$and $i_b^d,$ which
can be written as:

$$j_b^d=(n_b D_{bb}+n_s D_{bs})\frac{\partial n_b}{\partial x}+n_b
 D_{bs}\frac{\partial n_s}{\partial x}+(n_b K_{bb}+n_s K
_{bs})\frac{\partial n_b}{\partial \varphi }$$
$$i_b^d=(n_b F_{bb}+n_s F_{bs})\frac{\partial n_b}{\partial
\varphi }+n_b K_{bs}\frac{\partial n_s}{\partial x}+
(n_b K_{bb}+n_s K_{bs})\frac{\partial n_b}{\partial x} \eqno(7.24)$$

After routine calculations the coefficients $ D_{ik},$ $ F_{ik}$
and $ K_{ik}$ can be presented in following general form:

$$
D_{bb}={{\delta_b^2}\over{4}} \left ({T \over {\pi M_b}} \right)^{1 \over 2} 
J^D_{bb}, 
$$
$$
D_{bs}={{\delta_b\delta_s} \over {4 \sqrt{2\pi} }}
\left ({T\over {\mu_{bs}}}\right)^{1 \over 2} J^D_{bs},{~~}
D_{ik}=\{  D_{ik}, F_{ik}, K_{ik}\}
                                                       \eqno(7.25) 
$$

Here:

$$\mu _{ik}=\frac{M_iM_k}{M_i+M_k},i,k=s,b      \eqno(7.26)$$
where $M_s$ and $M_b$ are defined by formulae (2.16).

The expressions for $J_{ik}^{D}$ are presented in table1.It is easy
to see that the diffusion of breathers in real space ( x - space) is much
faster then the relaxation on $\varphi $.

\vskip 0.1 in
\begin{center}
{ACKNOWLEDGMENTS}
\end{center}
Authors are grateful for Prof. R.N.Gurgi and  Prof. L.N.Pastur for discussions.
Part of this work was supported by INTAS  grant Ref. No 94-3754. 
IVB and VGB are grateful for the hospitality of 
the Research Center of Crete/ FORTH.

\newpage
\begin{center}
{\bf References}\\
\end{center}
1. G. Zaslavskii and R.Z. Sagdeev,  Introduction to Nonlinear Physics
(Nauka, Moscow, 1988).\\ 
2. N.J. Zabusky and M.D. Kruskal, Phys. Rev. Lett. 15 , 240-243 (1965).\\ 
3. C.S. Gardner, J.M. Greene, M.D. Kruskal and R. Miura, 
Phys. Rev. Lett. 19, 1095-1097 (1967) \\ 
4. V.E. Zakharov, S.V. Manakov, S.P. Novikov and L.P. Pitaevskii,  
Theory of Solitons  (Moscow, Nauka, 1980) .\\
5. L.D. Faddeev and L.A. Tahtadzhan, Hamiltonian methods in the theory
of solitons  (Nauka, Moscow, 1986).\\ 
6. V.E. Zakharov, Zh.Eksp.Theor.Fiz. 60, 993 -1000 (1971).\\ 
7. T.R. Koehier,A.R. Bishop, J.A. Krumhansl, J.A. and J.R. Schrieffer,
Sol.State.Comm., 17, 1515-1519 (1975) \\
8. Y. Wada, J.R. Schrieffer, Phys.Rev. B 18, 3897-3912 (1978).\\
9. N. Theodorakopoulos, Z. Phys. B 33, 385-390 (1979).\\
10. K.Z. Fesser, Z. Phys.B 39, 47-52 (1979). \\
11. K. Sasaki and K. Maki, Phys. Rev.B 35, 257 - 262 (1987) \\
12. I.V. Baryakhtar, V.G. Baryakhtar and E.N. Economou,
Phys.Lett. A207, 67-71 (1995).\\
13. V.G.Baryakhtar,I.V.Baryakhtar,B.A.Ivanov and A.L.Sukstansky, in:
Proc. of the II Int. Symp. on Selected Topics in Statistical Mechanics,
JINR, Dubna,417-426 (1981).\\
14. O.M.Braun and Yu.S. Kivshar, Phys.Rep. 306,N1-2,1-108 (1998).
\newpage
\hspace{1.0in}

$Table 1$. Diffusion coefficients for breather \\
{\small
\hspace{1.0 in}
\begin{tabular}{|l|l|l|l|l|}
\hline
$ik$ &  \multicolumn{3}{c|}{$ T / \mu _{ik} \ \ll 1-\omega _0^2{~}$}
 & 
$ 1 \gg T/\mu _{ik}\gg 1-\omega _0^2{~}$   
\\ 
\hline
\hline
$\backslash $ & $\quad \quad {\it J}^{ D}_{ik}$ & 
${\it J}^{ K}_{ik} $ &
$  {\it J}^ { F} _{ik}   {~} $ &
$ {\it J} ^{ D} _{ik}$ 
\\ \hline
$bs$ & 
$[\ln \frac{1+\omega _0}{1-\omega _0}]^2$ & 
$\pi \ln\frac{1+\omega _0}{1-\omega _0}$ & 
$\quad {\pi^2}$ &
$[\ln \frac {2 \gamma \mu_{bs}}{T}]^2+C$  
\\ 
\hline
$bb$ & 
$[\ln \frac{\gamma \omega _0^2 M_b}{(1-\omega _0^2)T}]^2+C$ & 
$\pi \ln \frac{\gamma \omega _0^2 M_b}{(1-\omega _0^2)T}$ & 
$\quad {\pi ^2}$ &
$4 ([\ln\frac{\gamma M_b} {T}]^2+C)$  
\\ \hline
\end{tabular}
}

\end{document}